\documentclass[conference]{IEEEtran}

\IEEEoverridecommandlockouts

\usepackage[table]{xcolor}
\usepackage{url}
\usepackage{amssymb}
\usepackage{enumerate}
\usepackage{relsize}
\usepackage{url} 
\usepackage{multirow}
\usepackage{multicol}
\usepackage{rotating}
\usepackage{relsize}
\usepackage{bm}
\usepackage{graphicx}
\usepackage{hyperref}
\usepackage{multicol}
\usepackage{setspace}
\usepackage{enumerate}
\usepackage{footnote}
\usepackage{authblk}
\usepackage{float}
\usepackage{longtable}
\usepackage{tabularx}
\graphicspath{{Image/}}
\usepackage{etoolbox}
\usepackage[top=19mm,bottom=19mm,right=19mm,left=19mm]{geometry}
\usepackage{afterpage}
\usepackage{textcomp}

\begin{document}

\newgeometry{top=25mm,bottom=19mm,right=19mm,left=19mm}

\title{Data-Driven Exploration of Factors Affecting Federal Student Loan Repayment\\}

\author{Bin Luo, Qi Zhang, Somya D. Mohanty
\thanks{B. Luo is with the Department
of Mathematics and Statistics, University of North Carolina at Greensboro, Greensboro,
NC, 27412 USA e-mail: b\_luo@uncg.edu.}
\thanks{Q. Zhang is with the Department of Mathematics and Statistics, University of North Carolina at Greensboro, Greensboro,
NC, 27412 USA e-mail:q\_zhang@uncg.edu.}
\thanks{Somya D. Mohanty is with the Department of Computer Science, University of North Carolina at Greensboro, Greensboro,
NC, 27412 USA e-mail:sdmohant@uncg.edu.}}

\maketitle

\begin{abstract}
Student loans occupy a significant portion of the federal budget, as well as, the largest financial burden in terms of debt for graduates. This paper explores data-driven approaches towards understanding the repayment of such loans. Using statistical and machine learning models on the Collage Scorecard Data, this research focuses on extracting and identifying key factors affecting the repayment of a student loan. The specific factors can be used to develop models which provide predictive capability towards repayment rate, detect irregularities/non-repayment, and help understand the intricacies of student loans.

\end{abstract}



\section{Introduction}

In the United States, a large portion of college education is financed through federal student loans. Such loans reduce the financial burden enabling students to participate in higher education. In such cases, the lender could be a bank, credit union, the borrower’s school, and/or the Department of Education. The federal government issues \$1.45 trillion of student loan debt each year, and more than 90\% of student debt today is in the form of federal loans. In recent reports, currently there are approximately 44.2 million Americans with student loan debts \cite{frs}. 

Repayment of such loans includes a multitude of options, but most participate in a standard ten year plan with 120 equal monthly payments. With the federal government being one of the largest lenders in student loans, a key objective of the agency is to quantify the repayment capability of the borrower. A report from the federal reserve system states that individuals currently carry an average of over \$37,000 in unpaid loans, along with the average monthly student loan repayment at \$351 (for a borrower aged 20 to 30 years) \cite{frs}. With a reported \$31,099 as the 2016 nominal median income per capita \cite{cb}, the statistics illustrate a growing problem among college graduates in terms of loan repayment. Although loans cover the rising cost of higher education, subsequent debt can make it difficult to reap the benefits of the investment.

As expected, most college graduates carry student loan debt for much longer than they expected. The intuition behind incurring student loans, is to enable expansion of employment opportunities or movement towards a better position and/or job prospects, through higher education, where repayment can be done though additional income of an individual. However, students who are unable to complete their college education, completely counters this intuition. With the additional burden of college loans, and relatively similar income as before, individuals may face a key hurdle in becoming debt free. In comparison, repayment of student loans might not be an issue with individuals coming from a high income family. In such situations parents/family members may contribute to the loan and help alleviate some of the burden in the loan repayment.

These cases provide key questions towards analysis of the college debt environment. In the first case, a negative relationship might exist between one’s repayment rate to the education level or completion rate. Similarly, using the second example, correlations between repayment behaviour and the individual's family income can shed some insight into the repayment rates. Additional predictors of repayment rate are also discussed in Mezza and Sommer (2015)\cite{mezza}, Kelchen and Li (2017)\cite{kelchen}, Looney and Yannelis (2015)\cite{looney}.

This paper explores such questions and the factors which affect the repayment rate by using data driven analytics on the College Scorecard dataset \cite{data}. This rich dataset from the U.S. Department of Education provides records for student completion, debt and repayment, earnings, and many other key variables observed during 2007 - 2014. Apart from the aforementioned questions, the study also will examine relationships between latent variables in the dataset such as student demographics, college characteristics, geography, education level, etc. The paper also highlights key findings which can help enable better decision making for managing student loans by federal government and other lenders. 

\section{Background}

\subsection{Related work}
Prior research in this area utilized nationally representative datasets in order to study the predictors of student loans. In the work done by Mezza \textit{et. al.} (2015)\cite{mezza},  research was conducted using a dataset that anonymously combines credit bureau records, individual characteristics, and federal student loan information. It showed that borrower-level credit characteristics are strong predictors of student loan repayment, especially for young borrowers. They also found that individuals with lower educational levels are more likely to become delinquent in their loan repayment. A similar study by Looney and Yannelis (2015)\cite{looney} used administrative data from deidentified tax records to provide evidence that  high-income families have higher completion rate, low rates of unemployment and 
high earnings, which in turn leads to higher repayment rates of student loans. 

Other studies \cite{Hillman} have found that nonprofit or public four year colleges have higher repayment rate than for-profit and two-year colleges. Repayment rates may also be affected by geographic area, as institution located in rural areas\cite{Webber}. In the work done by Chen \textit{et. al.} \cite{Chen}, gender has also been found to be one of the factors. Different studies done in the domain report disparate factors associated with the repayment of student loans without any consensus among the reported results. The observations of inconsistencies of the factors in the studies can attributed to the different analytical methodologies, and the disparate datasets the methods were applied to. 

One of the first efforts to apply methods on a standardized dataset was in the research conducted by Kelchen and Li (2017) \cite{kelchen}, by using the College Scorecard data. In their study, the dataset was used to test associations between institutional characteristics and one year loan repayment rates. The study also aimed to analyze the effects on the repayment based ond the number of years after a student has left college. The goal of the study is to develop multiple linear regressions capable of analyzing repayment using predictor variables. While the study choose a fixed list of features (based on prior research) to be their predictor variables, a better model of establishing such variables would be to conduct an exhaustive search of all available features in the dataset. The research presented in the paper aims to identify such features using a multi-model approach, where the models are able to suggest the best performing predictors in the dataset. 

\subsection{Dataset Description}
\label{sec:datadec}
The College Scorecard dataset \cite{data} was created to enable comparative analysis of education cost for multiple universities / degree-granting colleges. The dataset includes financial aid and tax information data from 1996 to 2014 for all undergraduate institutions of higher education, this includes institution which receive federal financial aid. 

Towards the analysis of variables present in the dataset, our study focuses on subset of data from years 2007 to 2014, where the repayment rate data is provided. After data cleaning, there are 1859 variables included in the dataset, and are categorized as nine categories --- \textit{School, Academics, Admissions, Cost, Student, Completion, Aid, Earnings, }and \textit{Repayment}. `\textit{School}' --- basic descriptive information about the colleges; `academics' --- the types of academic offerings available at each institution; `\textit{admissions}' --- admissions rates and SAT/ACT scores of students; `\textit{cost}' --- information about the costs of tuitions and fees; `\textit{student}' --- demographic information about the student body of the institution; `\textit{aid}' --- information about financial aid; `\textit{completion}' --- different completion rates; `\textit{earnings}' --- the mean and median incomes; `\textit{repayment}' --- information about the burden of attending college and the loan performance for each institution \cite{dr}. Each group is further defined by the variables included in the category. Table \ref{tab:cat_var}, shows the number of variables in each of these categories.

\begin{table}
\caption{Number of variables in each category}
\centering
\begin{tabular}{cc}
\hline\hline
Category& Number of Variables\\
\hline
Repayment & 131\\
School&170\\
Academics&228\\
Admission&25\\
Cost&65\\
Student&96\\
Completion&1031\\
Aid&40\\
Earnings&73\\
\hline\hline
\end{tabular}
\label{tab:cat_var}
\end{table}

\subsection{Dataset Variables}
The dataset provides a college level overview of whether students are repaying their debts. With the goal of the study to analyze the repayment rate, we focus on the repayment rate observed for students who graduated between years of 2006 - 2011. The repayment rate category records the percentage of college graduated students who had federal loans and repaid their debt in --- one year ($RPY\_1YR\_RT$), three years ($RPY\_3YR\_RT$), five years ($RPY\_5YR\_RT$), and seven years ($RPY\_7YR\_RT$). The variables are further sub-categorized by the demographic variables, leading to a total of 131 variables.

\section{Methods}

\subsection{Data Preprocessing}
\label{sec:preprocess}
Analysis of the College Scorecard dataset was done with the merged data from 2007 to 2014. Data cleanup was performed to remove any features with text, and transformation was done for categorical variable into integer values. Observations or features with more than 70\% of missing data are removed. 
Unreadable data was replaced with zero (0) values in the dataset.

In order to reduce the dimensions of the dataset, we screened the features based on their variance and the correlation with repayment rate. Features with the lowest variance ($a\%$) were removed, for example, features containing where most of their observed values were 0 are removed. The top $b$ features with the highest correlation with the repayment rate were selected for analysis. Multiple combinations of $(a,b)$ were also generated as a part of the dataset to be used in machine learning models described in the following sections. 


\subsection{Principle Component Analysis and Linear Regression}
A key component in reducing data redundancy, is to identify patterns of correlation within the variables of the data. The identified highly correlated variables can be merged to form simpler representative variables, enabling to easier understanding of predictive models. As discussed previously (Section \ref{sec:datadec}), all the features can be divided into 8 categories: School, Admission, Academics, Student, Cost, Aid, Completion, Earnings. In these categories, features could be highly correlated, especially for those belonging to the same category. 

Principle Component Analysis(PCA) \cite{PCA}, is a suitable method to perform linear dimensionality reduction on such correlated features. PCA is a statistical procedure which converts a set of observations (for possibly correlated variables), into a set of values of linearly uncorrelated variables called principal components. This transformation is defined in such a way that the first principal component has the largest possible variance, and each succeeding component in turn has the highest variance under the key constraint, that it is orthogonal to the preceding components. However, if we apply PCA on the whole dataset (all features), the resulting principle component will be difficult to interpret. Therefore, we decided to apply PCA on each category of features individually, exploiting the latent similarities within the categories. The approach tries to balance the necessity of dimensionality reduction, while allowing for interpretation of the principal components. A cut-off value $c$ was used for explained variance, to determine the number of components for each category. 

The resulting transformed features were fitted to a linear regression model for repayment rate. The model was used for analyzing the repayment rate by evaluating the features, and performing hypotheses testing for the feature components from each category. The observation of repayment rate from the dataset was transformed to a log-based model where the repayment rate $p$ was replaced by $log(p/(1-p))$ when fitting the linear regression model. Then we have estimate $\hat{p}=\frac{\exp(\hat{y})}{1+\exp(\hat{y})}$, where $\hat{y}$ denotes the estimated transformed repayment rate.

\subsection{Elastic Net}
Apart form using PCA to reduce the dimensionality on the dataset, Elastic Net \cite{EN} algorithm was also used for variable selection, and coefficient estimation for highly correlated features. Elastic net is a penalized regression method, which linearly combines both $L_1$ and $L_2$ penalties of the Lasso \cite{LASSO} and Ridge \cite{Ridge} methods. The estimator from elastic net is given by:

\begin{equation}
    \hat{\beta}=\arg min (\frac{1}{2n}|| Y-X\beta||_2^2+\lambda \alpha|\beta|+\frac{1}{2}\lambda (1-\alpha) ||\beta_2||_2^2)
\end{equation}

At $\alpha=1$ and $\alpha=0$, the algorithm operates as Lasso and Ridge estimator respectively. The Lasso estimator is capable of selecting variables and estimating coefficients simultaneously. However, it has a limitation that, when a group of highly correlated variables exist, the Lasso tends to select a single variable from that group and ignore the others. The elastic net was designed to overcome this limitation by incorporating an additional $L2$ penalty. With the advantages of the Elastic Net over the regular Lasso model, the algorithm was used for identifying important features which affect repayment rate.

\subsection{Random Forest}
While the assumption of the underlying model (in prediction of repayment rate) to be linear for regression methods might be correct, the study also explored utilization of non-linear models such as Random Forest \cite{RF}. Random forests is an ensemble learning method, that constructs a multitude of decision trees \cite{DT} during training to fit the data, and can have a categorical or a continuous value output. For example, the model is capable of supervised learning with prediction of classes of outcome, and similarly can output prediction values (mean values using regression) from the observations. Furthermore, the algorithm is also capable of identifying important features which influence the decision of the models. Thus, prediction models based on Random Forest were developed in the study.

\section{Analysis and Results}
The pre-processing on the original dataset lead to the threshold calculation of top $a$ features as $a=10$, and top $b$ features as $b=500$ on multiple combinations of $(a,b)$. The output dataset for analysis consisted of of 500 features and 35,027 observations. The first step in analysis of the dataset for features affecting student loan repayment was the period of repayment variable (based on the factors considered in prior literature \cite{kelchen}).

\begin{figure}
  	\centering
  	\includegraphics[width=.5\textwidth]{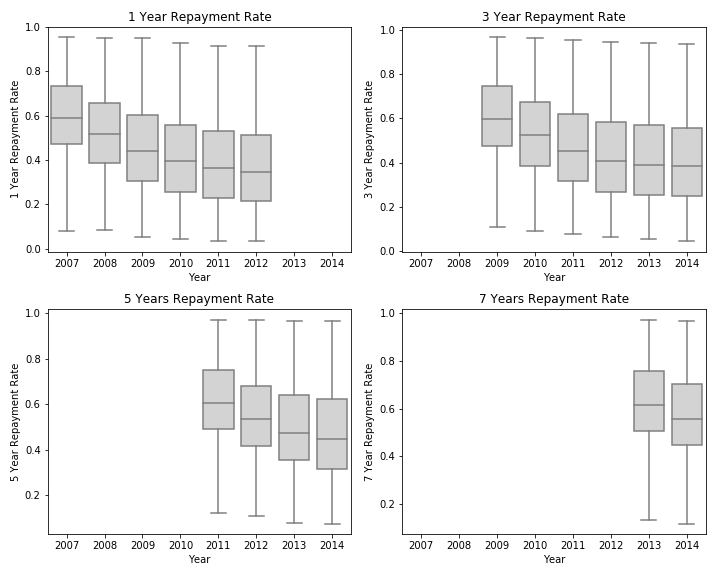}
  	\caption{Box-Plot of repayment rates - In different years}
  	\label{fig:rep_rate}
  \end{figure}

Figure \ref{fig:rep_rate}, shows the distribution of different years of repayment rates. The rates have been normalized to be in the range of 0 to 1. The median repayment rates are observed to be in the range of 0.4 to 0.6. An overall trend across all of the different categories (number of years) is the gradual decline in loan repayment, indicating fewer borrowers are repaying their debt in recent years. 

  \begin{figure}
  	\centering
  	\includegraphics[width=.4\textwidth]{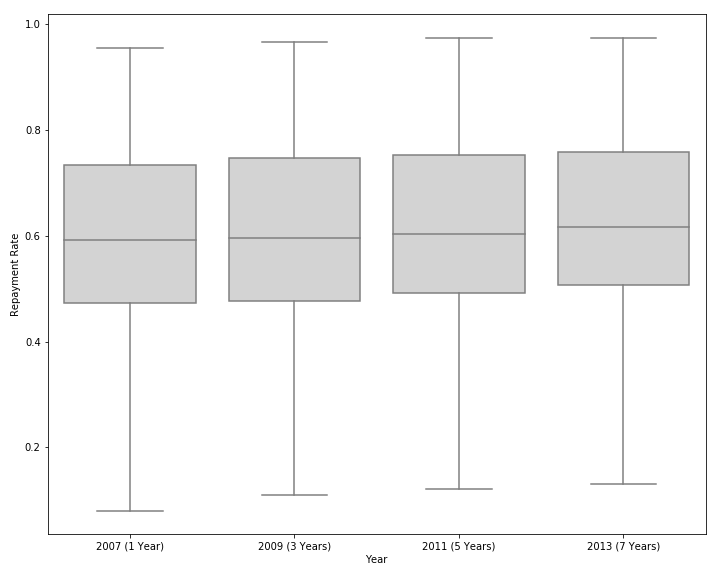}
  	 \caption{Box-Plot of different time periods repayment rate - Students who graduated in 2006}
  	 \label{fig:rep_2006}
  \end{figure}

Figure \ref{fig:rep_2006}, describes the repayment rate for different time periods (1 year, 3 year, 5 year, and 7 year) for students who graduated in 2006. More specifically, the one year repayment rate for the 2006 batch was collected in 2007, for 3 year it was collected in 2009, and so on. Analyzing the repayment rate across the periods shows no significant deviation in repayment behaviour, with a median rate across the time periods being about 0.6. The study was also conducted across the batch of 2007 and 2008, resulting in similar results. Due to the similar repayment rate for different time periods and fewer missing value for 1-year repayment rate, the study choose to proceed with the analysis using 1-year repayment rate only.
%



%



\subsection{Potential Predictors} 

  \begin{figure}
  	\centering{
  	\includegraphics[width=.5\textwidth]{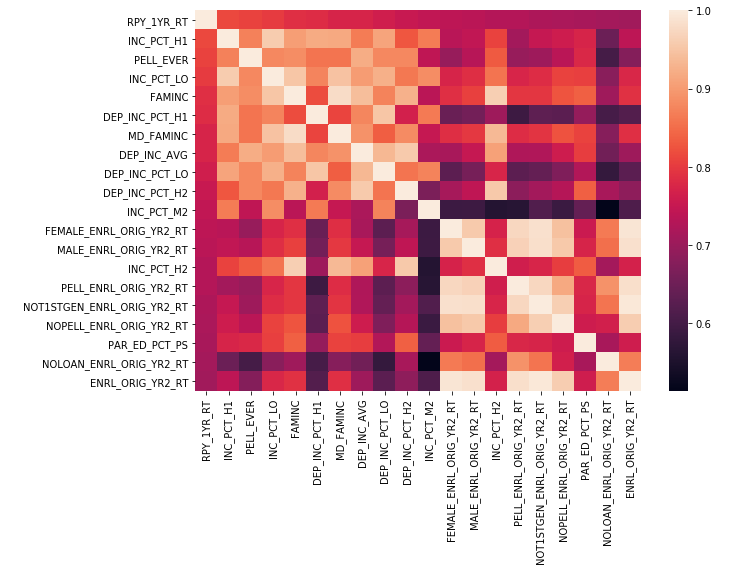}
  	 \caption{Feature Correlation Heatmap - Top 20 Highly Correlated Variables}
  	 \label{fig:corr}
  	 }
  \end{figure}

In order to find the independent variables, the absolute values of correlation were ranked, and then filtered for highly correlated variables. Using the correlation heatmap (Figure \ref{fig:corr}), from the top 20 highly correlated variables, we can identify that most variables emerge from categories of student (family income) and degree completion. Also within a single category, multiple variables represent very similar features. For example, while both $INC\_PCT\_LO$ and $DEP\_INC\_PCT\_LO$ indicate the percentage of aided students whose family income is between \$0-\$30,000, but latter ($DEP\_INC\_PCT\_LO$) is for students who are financially independent. In other words, the population targeted by the latter is a sub-group of the former. Variables are further correlated based on the sibling aspect, where one variable is a branch of another. 

As mentioned in Section \ref{sec:preprocess}, the complete dataset contains 1859 variables. In the development of regression models large dimensionality of variables can result in a highly complex system. With the aforementioned relationships between the variables, dimensionality reduction methods can be used for reducing the complexity of models. 

After the preprocessing of dataset, the remaining remaining features for each category are --- \textit{School} - 9, \textit{Admission} - 20, \textit{Academics} - 33, \textit{Student} - 40, \textit{Cost} - 20, \textit{Aid} - 15, '\textit{Completion} - 317, and \textit{Earnings} - 46. As a result, the pre-screening process reduced the number of features from 1859 features/variables (in the original dataset) to 500 variables used in further analysis.

\begin{figure}[h]
	\centering
	\includegraphics[width=.5\textwidth]{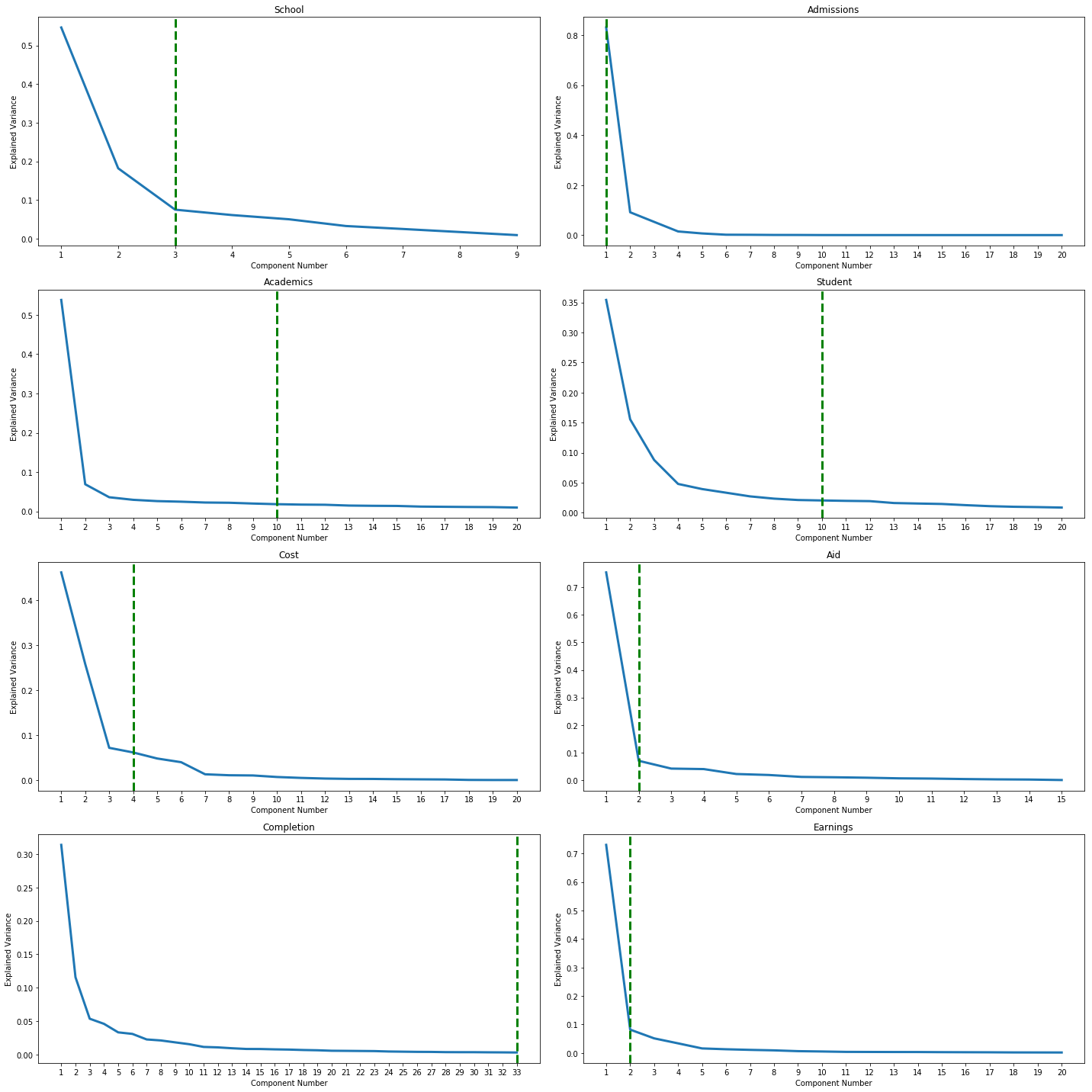}
	\caption{Scree Plot for Variance - Feature Category: The green dash line marks the minimum number of components $PC_{i}^{c}$ which explain at least 80\% of variance within each category.}
	\label{fig:PCA}
\end{figure}

Using Principal Component Analysis (PCA) on the remaining features in each category, the features were transformed to $n_k$ principal components - $PC_{i = 1 \cdots n_k}^{k}$, where, for $kth$ category of the features, $i$ ranges from $1$ - $n_k$, and $n_k$ being the number of features in the category. The variance in each of the principal components was plotted using a Scree plot (shown in Figure \ref{fig:PCA}) for each category. It can be observed that the $PC_{i = 1 \cdots n_k}^{k}$ in most categories, --- \textit{School}, \textit{Admission}, \textit{Cost}, \textit{Aid}, and \textit{Earnings}, needs only a few components ($i < 10$) to cover up to 80\% \footnote{80\% threshold was chosen as a first intuitive measure, and later cross-validated using regression methods} of variance in features. However in the category of \textit{Completion}, the threshold variance could not be captured with low number of variables, describing inconsistency among the features in the category. The PCA analysis resulted in further reduction of variable size from 500 (pre-screening) to 65 (after PCA).

The dataset with the reduced dimensions (65 variables from 80\% threshold cutoff) was fit on to a linear regression model. The model resulted in the R-Squared measure of $R^2=0.842$, indicating approximately 84\% of variance on transformed repayment rate (dependent variable $Y$) can be explained by the linear relationship between the reduced components (independent variables $X$) chosen by the 80\% threshold of PCA. The root mean square error ($RMSE$) of the model is $RMSE=0.0197$ ($2\%$ estimated standard deviation in prediction error) observed over 10-fold cross validation on the dataset.

\begin{figure}[h]
	\centering
	\includegraphics[width=.5\textwidth]{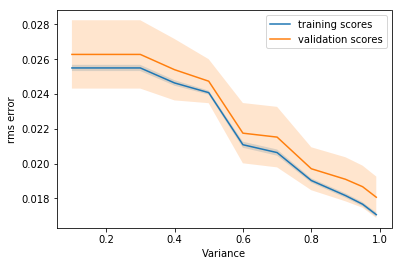}
	\caption{Train / Validation Curve - $RMSE$ vs. Variance.}
	\label{fig:validation_cruve}
\end{figure} 

\begin{figure}[ht!]
 	\centering
 	\includegraphics[width=.5\textwidth]{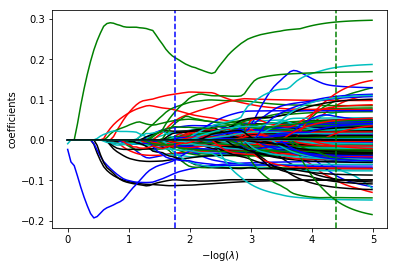}
 	\caption{Elastic Net solution paths - Coefficients vs. $-\log(\lambda)$: The green dash line and blue dash line represent the optimal tuning parameter and solution chosen by cross validation, for $\frac{\lambda_{min}}{\lambda_{max}}= 5\mathrm{e}{-6}$ and $5\mathrm{e}{-6}$, respectively.}
 	 	\label{fig:ENP}
 \end{figure}

\begin{table*}[!ht]
\centering
\caption{Important Features - Elastic Net.}
\label{tb:ENT}
\scalebox{0.65}{
\begin{tabular}{|l|p{22cm}|c|}
\hline
\textbf{\#}& \textbf{Feature description and Name}                                                                                                                & \textbf{Category} \\ \hline \hline
1. & Control of institution $CONTROL$                                                                                                       & school \\\hline
2. & Total enrollment (share) Africa-American students $UGDS\_BLACK$                                                                        & student \\\hline
3. & Total enrollment (share) White students (non-Hispanic) $UGDS\_WHITENH$                                                                 & student\\\hline
4. & \% of students receiving Pell Grant $PCTPELL$                                                                                          & aid\\ \hline
5. & \% Withdrawn within 2 years $WDRAW\_ORIG\_YR2\_RT$                                                                                     & completion\\\hline
6. & \% Transferred to a 2-year institution and withdrew within 2 years $WDRAW\_2YR\_TRANS\_YR2\_RT$                                        & completion\\\hline
7. & \% Student never received federal loan and withdrew  $NOLOAN\_WDRAW\_ORIG\_YR2\_RT$                                                    & completion\\\hline
8. & \% Not-first-generation students transferred to a 4-year institution and withdrew (2-year) $NOT1STGEN\_WDRAW\_4YR\_TRANS\_YR2\_RT$     & completion\\\hline
9. & \% Not-first-generation students transferred to a 2-year institution and withdrew (2-year) $NOT1STGEN\_WDRAW\_2YR\_TRANS\_YR2\_RT$     & completion\\\hline
10. & \% who transferred to 2-year institution and withdrew within 3 years $WDRAW\_2YR\_TRANS\_YR3\_RT$                                     & completion\\\hline
11. & \% students never received federal loan and withdrew within 3-years $NOLOAN\_WDRAW\_ORIG\_YR3\_RT$                                    & completion\\\hline
12. & \% students never received federal loan and transferred 4-year institution and withdrew (3-year) $NOLOAN\_WDRAW\_4YR\_TRANS\_YR3\_RT$ & completion\\\hline
13. & \% Not-first-generation students transferred to a 2-year institution and withdrew (3-year) $NOT1STGEN\_WDRAW\_2YR\_TRANS\_YR3\_RT$    & completion\\\hline
14. & \% students transferred to 2-year institution and withdrew (4-year) $WDRAW\_2YR\_TRANS\_YR4\_RT$                                      & completion\\\hline
15. & \% students never received federal loan and transferred 4-year institution and withdrew $NOLOAN\_WDRAW\_ORIG\_YR4\_RT$                & completion\\\hline
16. & \% students never received federal loan and transferred 2-year institution and withdrew $NOLOAN\_WDRAW\_2YR\_TRANS\_YR4\_RT$          & completion\\\hline
17. & \% aided students whose family income is \$0-30,000  $INC\_PCT\_LO$                                                                   & student\\\hline
18. & \% students who are financially dependent with family incomes \$0-30,000  $DEP\_INC\_PCT\_LO$                                     & student\\\hline
19. & \% first-generation students $PAR\_ED\_PCT\_1STGEN$                                                                                   & student\\\hline
20. & Aided students with family incomes \$48,001-75,000 $INC\_PCT\_M2$                                                                     & student\\\hline
21. & Aided students with family incomes between \$75,001-75,001-110,000 $INC\_PCT\_H1$                                                     & student\\\hline
22. & \% students whose parents' highest educational level is high school $PAR\_ED\_PCT\_HS$                                                & student\\\hline
23. & Average family income of dependent students in real 2015 dollars $DEP\_INC\_AVG$                                                      & student\\\hline
24. & Share of students who received a Pell Grant while in school $PELL\_EVER$                                                              & student\\\hline
25. & Share of first-generation students    $FIRST\_GEN$                                                                                    & student\\\hline
26. & Average family income in real 2015 dollars $FAMINC$                                                                                   & student\\ \hline
\end{tabular}
}
\label{tab:elastic}
\end{table*}

Using the prediction pipeline of the linear regression model, multiple cutoff threshold values for the variance in the components were evaluated. Apart from the 80\% ($.8$) threshold, features obtained $[0.1,0.2,0.3,0.4,0.5,0.6,0.7,0.9,0.95,0.99]$ array of cutoff values were used, and the linear model refitted to the features. Figure \ref{fig:validation_cruve}, shows the downward trend of validation curve ($RMSE$ value) with increasing threshold values of variance. The pattern was observed for both training and validation scores, suggesting increasing the number of features in the linear model from the initial assumption of 80\% threshold cutoff. 


While, the analysis suggests inclusion of all 500 features from the dataset, the resulting gain accuracy/performance of predictive models is negligible ($RMSE = 0.019$ at variance .8 - 65 features versus $RMSE = 0.018$ at variance 1 - all features). The study suggests limiting the model features to the 65 previously identified features, which can be used to best explain the predictive models and have the highest influence. Lower than .8 variance leads to heavy degradation of the linear models, and $>$ .8 variance for our features leads to more complex models with diminishing gains in predictive capability.


 In order to better understand the dependence of features on predictability, analysis was performed using Elastic-Net algorithms as described in the methodology. Cross-validation was conducted for tuning parameter selection under a fine grid of($\alpha, \lambda$), where $\alpha=[0.1,0.3,0.5,0.7,0.9]$ and $\frac{\lambda_{min}}{\lambda_{max}}=\gamma$. Figure \ref{fig:ENP}, shows the solution paths for the Elastic-Net model with $\alpha=0.9$. The blue dash line represents the optimal solution with $\gamma=5\mathrm{e}{-6}$, utilizing 432 features (from 500 total features). The resulting model reported an $RMSE$ score of $0.0177$, obtained by 10-fold cross validation. While the default model results in a very high accuracy in predicting the repayment rate, the complexity of large number of features can be reduced choosing higher value of $\gamma$. By setting $\gamma=5\mathrm{e}{-3}$, the optimal model (shown by green line) yields the RMSE score of $0.0223$ with the number of relevant features reduced to only 26 from 432.

\begin{table}[!ht]
	\begin{center}
	 \caption{RMSE - PCA-LS, Elastic-Net and Random-Forest: Computed by 10-fold Cross Validation}\label{table:RMSE}
	\begin{tabular}{lrrr}\\ \hline\hline
	    Method  & PCA-LS& Elastic Net & Random Forest \\ \hline
	    RMSE &   0.0197 & 0.0177 & 0.0153\\
	    \hline\hline
	\end{tabular}
	\end{center}
	\label{tab:RMSE_comp}
\end{table}

Table \ref{tab:elastic}, describes the 26 features observed from the second Elastic-Net model. Among the identified features, most of the selected features are from the overall categories \textit{'Student'} and \textit{'Completion'}. In the category of \textit{'Student'}, the relevant features are related to the family income of the student (percent of student), which is in turn consistent with the finding from the correlation analysis done previously. However, the selected features from category \textit{'Completion'} mainly focuses on the withdrawal rate for certain group of students. In comparison, the correlation analysis in the category suggested enrollment rate to be a key factor. The Elastic-Net model was also able to identify the categories of \textit{'Aid'} and \textit{'School'} to be relevant categories, which the correlation analysis was unable to report.

\begin{table*}[!ht]
\centering
\caption{Important Features - Random Forest}
\label{tb:RFT}
\scalebox{0.65}{
\begin{tabularx}{\textwidth}{|l|X|c|}
\hline
\textbf{\#}& \textbf{Feature description and Name}                                                                                                                & \textbf{Category} \\ \hline \hline
1. & Total enrollment (share) Africa-American students $UGDS\_BLACK$                                                                        & student \\\hline
2. & Total enrollment (share) White students (non-Hispanic) $UGDS\_WHITENH$                                                                 & student\\\hline
3. & \% of students receiving Pell Grant $PCTPELL$                                                                                          & aid\\ \hline
4. & \% Withdrawn within 2 years $WDRAW\_ORIG\_YR2\_RT$                                                                                     & completion\\\hline
5. & \% students received a Pell Grant enrolled within 2 years  $PELL\_ENRL\_ORIG\_YR2\_RT$                                                                                     & completion\\\hline
6. & \% aided students whose family income is \$0-30,000  $INC\_PCT\_LO$                                                                   & student\\\hline

7. & The median debt for female students $FEMALE\_DEBT\_MDN$                                                                   & aid\\\hline
8. & Share of students who received a Pell Grant while in school $PELL\_EVER$                                                              & student\\\hline
9. & Average family income in real 2015 dollars $FAMINC$                                                                                   & student\\ \hline
10. & Median family income in real 2015 dollars $MD\_FAMINC$                                                                                   & student\\ \hline
\end{tabularx}
}
\label{tab:rand_fea}
\end{table*}


\begin{figure}[h]
	\centering
	\label{fig:imfeature}
	\includegraphics[width=.5\textwidth]{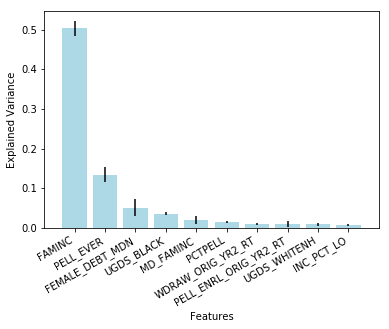}
	\caption{Feature importance - Random Forest: The blue bars are the feature importance of the forest, along with their inter-trees variability.}
	\label{fig:rand_fea}
\end{figure}

Expanding the analysis further, we also fitted a Random Forest Regression model (20 estimators/decision trees) to our dataset. Using 10-fold cross validation for the model, the observed $RMSE$ score for the model was $0.0153$. The reported score outperforms the best models obtained from Linear regression and Elastic-Net algorithms (Table \ref{tab:RMSE_comp}). The resulting Random Forest model was analyzed for feature importance, where features was analyzed for corresponding variance (of error) in predicting repayment rate. 

Figure \ref{fig:rand_fea}, shows the importance of features in the developed model. The 10 most important features observed in the model are outlined in Table \ref{tab:rand_fea}. The observed features align heavily with the identified features from Elastic-Net, where most of the chosen features were from the categories of \textit{'Student'} and \textit{'Completion'}. The feature with the highest importance observed is the Family Income ($FAMINC$) from the \textit{'Student'} category, suggesting the low loan default for students who have consistent income in their family. The observation suggests payment of the loan by their family members, which was also suggested by prior studies. The model also further emphasizes the importance of \textit{'Aid'} category for students being key category, where if a student has received a Pell Grant ($PCTPELL$), it weighs heavily (second most important feature) on their loan repayment. Other observed features in the model, suggest withdrawal rate ($WDRAW\_ORIG\_YR2\_RT$), and demographic ($UGDS\_BLACK$, $UGDS\_WHITENH$) / gender ($FEMALE\_DEBT\_MDN$) features contributing to the model.


\section{Conclusion and Future Work}
With the goal of the study to conduct a data-driven exploration of factors contributing to repayment rate of federal loans, our analysis determines the feature categories of \textit{'Student'}, \textit{'Completion'}, and \textit{'Aid'} being the key contributors to the repayment rate. Machine-learned models, as developed in our study can be used prediction of repayment rate, where Random Forest regression out performs Principal Component based Linear Regression, and Elastic-Net models. While the performance of the models are different, they consistently agree that the data features of student family income; withdrawing and enrolling rate; and Pell grant aid, are important factors to affect the repayment rate.

Future work in the area can include improved analysis of the data in the under performing models. For example, in Elastic-Net model Bayesian information criterion (BIC) \cite{BIC} can be used for tuning feature selection. For variable selection, we can also consider the minimax concave penalty (MCP)\cite{MCP} and the Smoothly Clipped Absolute Deviation (SCAD) penalty \cite{SCAD}, since they both possess better theoretical property than the Lasso penalty. Unsupervised learning can also be used to draw connections between feature groupings, and modeling features to discover key characteristics. Parameter tuning for models can also be explored to develop even more accurate predictive models.


\begin{thebibliography}{acmtrans}
\bibitem{SCAD}
Variable selection via nonconcave penalized likelihood and its oracle properties
\end{thebibliography}
\end{document}